\documentclass[]{article}

\usepackage{algorithm}
\usepackage{algorithmic}
\usepackage{amsmath}
\usepackage{fullpage}
\usepackage{graphicx}
\usepackage{multicol}
\usepackage{multirow}

\usepackage{rotating}

\usepackage{hyperref}

\newcommand{\A}{\mbox{$\mathsf{A}$}}

\newcommand{\CA}{\mbox{$\mathsf{CA}$}}
\newcommand{\LA}{\mbox{$\mathsf{LA}$}}

\newcommand{\mainalgname}{\textsc{LocAG}}
\newcommand{\firststagenum}[1]{{\mbox{FS#1}}}
\newcommand{\nonlocDSname}{NonLoc}

\providecommand{\keywords}[1]{\textbf{\textit{Keywords:}} #1}

\newcommand{\dept}{Department of Electrical Engineering \& Computer Science, United States Military Academy,
West Point, NY, 10996, U.S.A.}

\title{Faster Location in Combinatorial Interaction Testing}

\author{Ryan E. Dougherty\thanks{\dept\ Email: \texttt{ryan.dougherty@westpoint.edu}}
\and Dylan N. Green\thanks{\dept\ Email: \texttt{dylan.green@westpoint.edu}}
\and Grace M. Kim\thanks{\dept\ Email: \texttt{grace.kim@westpoint.edu}}
}

\begin{document}

\maketitle

\begin{abstract}
Factors within a large-scale software system that simultaneously interact and strongly impact the system's response under a configuration are often difficult to identify.
Although screening such a system for the existence of such interactions is important, determining their location is more useful for system engineers. 
Combinatorial interaction testing (CIT) concerns creation of test suites that nonadaptively either detect or locate the desired interactions, each of at most a specified size or show that no such set exists.

Under the assumption that there are at most a given number of such interactions causing such a response, 
locating arrays (\LA s) guarantee unique location for every such set of interactions and an algorithm to deal with outliers and nondeterministic behavior from real systems, we additionally require the \LA s to have a ``separation'' between these collections. 
State-of-the-art approaches generate \LA s that can locate at most one interaction of size at most three, due to the massive number of interaction combinations for larger parameters if no constraints are given. 
This paper presents \mainalgname, a two-stage algorithm that generates (unconstrained) \LA s using a simple, but powerful partitioning strategy of these combinations.
In particular, we are able to generate \LA s with more factors, with any desired separation, and greater interaction size than existing approaches.
\end{abstract}

\keywords{locating array, covering array, genetic algorithm}

\section{Introduction}

The analysis of large-scale complex systems has always been a challenge, and is necessary as such systems are ubiquitous.
One difficult aspect of this analysis is ensuring correct operation, which is accomplished often with testing. 
If the number of system configurations is sufficiently high, then exhaustive testing may be infeasible. 
Combinatorial interaction testing (CIT) involves designing and analyzing a test suite to cover many interactions of system components within the tests. 
One of the most central objects in CIT is a covering array, which is a formal model that guarantees all interactions of size at most a specified number $t$ appear at least a specified number of times $\lambda$.
Utilization of a covering array within a testing scenario is simple: for each of the covering array's rows (i.e., a single test), set the corresponding system component to the specified input.
The execution of each test yields a response from the system. 
A covering array guarantees that if the response is desired from each of the test runs, then any undesired behavior in the system is the result of more than $t$ components. 
The limitation of covering arrays is that they do not provide \emph{recovery} of what interactions caused the response or what settings to components significantly impact the response. 

Localizing the fault(s) within a system is a well attacked problem; one strategy for CIT is to run a set of tests and note which tests passed and failed. 
Second, test each possible interaction that appears within each failed test. 
Third, estimate the likelihood that the interaction did indeed caused the fault.
Finally, generate new tests based on this likelihood. 
This approach is known as \emph{adaptive} testing; however, some testing environments do not support such a strategy, such as any environment that is nondeterministic (either in simulation or measurement).
Locating arrays, introduced by Colbourn and McClary \cite{colbourn_locating_2008}, are a nonadaptive testing strategy that in addition provide a guarantee that a set of interactions of a given size $d$ correspond to a unique set of tests.
Therefore, the interactions responsible for the response can (in principle) be recovered.
% Determining the number $d$ correctly in advance remains a challenge; to address this, we address the locating array variant of uniquely recovering any set of interactions of size \emph{at most} $d$.

Despite the general framework, the constructions of locating arrays have only been well studied for $d=1$, whether mathematical \cite{colbourn_locating_2008} or computational \cite{konishi_finding_2020,konishi_using_2020}.
This restriction is mostly due to the computational cost involved (see Section~\ref{sec:the_two_stage_approach}) or because of practical motivation, such as systems with ``heavy-hitter'' assumptions \cite{seidel_robustness_2019}.

In this work, we present an algorithm \mainalgname\ that computationally constructs locating arrays much faster than existing approaches.
One consequence is that we can construct locating arrays for larger and more complex systems than these approaches, including locating at most $d = 2$ interactions for some systems.
Our algorithm contains two stages; the first uses a simple but powerful partitioning strategy, and the second uses a genetic algorithm. 
As far as we are aware, this work is the first to produce non-trivial locating arrays with $d = 2$.
Additionally, our formulation allows for redundancy in the underlying models in cases of testing environment nondeterminism. 
The paper is organized as follows:
\begin{itemize}
    \item Section~\ref{sec:preliminaries} includes preliminaries about CIT and genetic algorithms.
    \item Section~\ref{sec:the_two_stage_approach} includes details about, and pseudocode for, our two-stage approach.
    \item Section~\ref{sec:experimental_evaluation} contains our experimental design, main research questions, and experimental evaluations.
    \item Section~\ref{sec:threats} contains threats to validity about our method.
    \item Section~\ref{sec:related_work} contains related work about locating arrays and applications of evolutionary computation to CIT.
    \item Section~\ref{sec:future_work} provides future work directions.
    \item Section~\ref{sec:conclusion} concludes the paper.
\end{itemize}

\section{Preliminaries}\label{sec:preliminaries}

\subsection{CIT}

We develop a formal testing model for CIT.
Let $F_1, \cdots, F_k$ be $k$ \emph{factors}, and each factor $F_i$ has an associated set of $v$ \emph{levels} $\{x_{i,1}, \cdots, x_{i,v}\}$.
A \emph{test} is an assignment to $F_i$ one of its levels, for each $1 \le i \le k$.
An \emph{interaction} is a set of pairs $(F_i, x_i)$ where $x_i$ is a level for $F_i$; if an interaction has size $t$, it is a \emph{$t$-way interaction}.
Here, $t$ is the \emph{strength} of the interaction.
Let $\A$ be an array with $N$ rows and $k$ columns, where each column $i$ corresponds to factor $F_i$ and each entry in column $i$ is one of the levels for $F_i$.
Let $T$ be a $t$-way interaction, and $\rho_{\A}(T)$ to be the set of rows in $\A$ for which $T$ appears.
The formal definition of ``appears'' within a row $r$ is that for each column-value pair $(F_i, x_i)$ within $T$, row $r$ has the level for factor $F_i$ set to $x_i$.
If $T$ appears in a row $r$, then we say $T$ is \emph{covered} in $r$.
Suppose that every $t$-way interaction $T$ has the property that $\rho_{\A}(T)$ is nonempty (i.e., $|\rho_{\A}(T)| \ge 1$); then we denote $\A$ as a \emph{covering array} $\CA(N; t, k, v)$. 

Much research has been devoted to finding the smallest covering array for which the coverage property still holds, i.e., the smallest $N$ for which a $\CA(N; t, k, v)$ exists. 
This smallest $N$ is the \emph{covering array number} \cite{colbourn_combinatorial_2004}, and corresponds to the minimum number of tests needed for such a system. 
Empirical analysis on real-world systems shows that pair-wise testing ($t=2$) can detect approximately half of software faults, whereas higher-strength testing ($t=6$) can detect over 99\% of faults \cite{colbourn_combinatorial_2004}.
Therefore, methods to construct covering arrays of higher-strength are desired. 

Figure~\ref{fig:ca_example} gives an example of a covering array $\CA(6; 2,4,2)$.
It has $N=6$ rows, $k=4$ columns, $v=2$ different entries (0 and 1), and for every $t=2$ columns, all $v^t = 4$ combinations appear at least once.
As an example, for columns 1 and 3 (indexing from 1), we compute $\rho_{\A}(T)$ for all interactions $T$ in these columns:

\begin{itemize}
    \item When $T = \{(1,0), (3,0)\}$, $\rho_{\A}(T) = \{1, 6\}$.
    \item When $T = \{(1,0), (3,1)\}$, $\rho_{\A}(T) = \{2\}$.
    \item When $T = \{(1,1), (3,0)\}$, $\rho_{\A}(T) = \{4, 5\}$.
    \item When $T = \{(1,1), (3,1)\}$, $\rho_{\A}(T) = \{3\}$.
\end{itemize}

\begin{figure}
    \centering
\[\begin{array}{cccc}
0 & 0 & 0 & 0 \\
0 & 1 & 1 & 1 \\
1 & 0 & 1 & 0 \\
1 & 1 & 0 & 1 \\
1 & 0 & 0 & 1 \\
0 & 1 & 0 & 0 
\end{array}\]

    \caption{A $\CA(6; 2, 4, 2)$.}
    \label{fig:ca_example}
\end{figure}

Of course, all of the other $\rho_{\A}(T)$ sets need to be verified for this array to be called a covering array. 
For two of the four example interactions, the second and fourth, there was only one test that witnessed the interaction.
If there is experimental noise or measurement error in the system, and if $\rho_{\A}(T)$ is ``small'' for an interaction $T$, then running the same test in which $T$ appears may yield different responses. 
On the other hand, if $\rho_{\A}(T)$ is ``large'', then the likelihood of a response related to $T$ being incorrect from the ground truth is small. 
Therefore, we impose another requirement on covering arrays, in that $|\rho_{\A}(T)| \ge \lambda$ for all interactions $T$ where $\lambda$ is a provided integer.
We denote such an array as $\CA_\lambda(N; t, k, v)$; here, $\lambda$ is the \emph{redundancy}.
For example, the covering array in Figure~\ref{fig:ca_example} has $\lambda = 1$ as all interactions are covered at least once.

A limitation of covering arrays is that they do not permit the identification of interactions causing the response; locating arrays are an extension that allows identification.
We first extend $\rho$ to include sets of interactions in a straightforward way.
Let $\mathcal{T}$ be a set of $d$ $t$-way interactions, where $d \ge 1$ is another provided integer.
Let $\A$ be a covering array. 
Define $\rho_{\A}(\mathcal{T})$ to be the set of rows any interaction in $\mathcal{T}$ appears in $\A$, i.e., $\rho_{\A}(T) = \bigcup_{T \in \mathcal{T}} \rho_{\A}(T)$.
Let $\mathcal{T}_1, \mathcal{T}_2$ be arbitrary distinct $d$-sets of $t$-way interactions. 
Suppose it is the case that the symmetric difference of $\rho_{\A}(\mathcal{T}_1), \rho_{\A}(\mathcal{T}_2)$ is always larger than $\lambda$; in other words, $\mathcal{T}_1 = \mathcal{T}_2$ if and only if
\[
    |(\rho_{\A}(\mathcal{T}_1) \cup \rho_{\A}(\mathcal{T}_2)) \setminus ((\rho_{\A}(\mathcal{T}_1) \cap \rho_{\A}(\mathcal{T}_2))| < \lambda.
\]
If so, then we denote $\A$ as a \emph{$(d,t,\lambda)$-locating array}, written as $\LA_\lambda(N; d, t, k, v)$.
When $\lambda=1$, we instead write $\A$ as a $(d,t)$-locating array.
When this symmetric-difference-on-rows property holds for $\mathcal{T}_1, \mathcal{T}_2$, we say that these two sets are a \emph{locating pair}, and a \emph{nonlocating pair} otherwise. 

An example of a locating array $\LA_{1}(7; 1, 2, 4, 2)$ is shown in Figure~\ref{fig:la_example}.
Since $d=1$, it is sufficient to determine that for any single interaction $T_1$ that the rows in which it appears is different than any other single interaction $T_2$.
We provide an example for verification. 
Suppose $T_1 = \{(1,0), (3,0)\}$ and $T_2 = \{(2,1), (4,0)\}$ are two interactions (again indexing columns from 1).
Then $\rho_{\A}(T_1) = \{4,5\}$ and $\rho_{\A}(T_2) = \{1, 2\}$.
Since the symmetric difference of $\{4,5\}$ and $\{1,2\}$ is 4, which is at least $\lambda=1$, we have that $T_1, T_2$ are a locating pair.
Verification would have to in principle examine every other pair of interactions.

\begin{figure}
    \centering
\[ \begin{array}{cccc}
0 & 1 & 1 & 0 \\
1 & 1 & 0 & 0 \\
1 & 0 & 0 & 1 \\
0 & 0 & 0 & 0 \\
0 & 1 & 0 & 1 \\
0 & 0 & 1 & 1 \\
1 & 0 & 1 & 0 \end{array}\] 

    \caption{A $\LA_{1}(7; 1, 2, 4, 2)$.}
    \label{fig:la_example}
\end{figure}

There are variants of the definitions that allow for sets of size \emph{at most} $d$ (denoted $\overline{d}$ instead of $d$), or individual interaction size \emph{at most} $t$ (denoted $\hat{t}$ instead of $t$), or no interaction contained within another (denoted $\overline{t}$ instead of $t$).
The motivation for these variants is that it is impractical to a priori determine the exact number of faulty interactions, or how large any interaction itself is.
In this work we focus strictly on building $(\overline{d},t)$-locating arrays as the other variants are simple modifications of the underlying method.
For similar reasons, we assume that all factors have the same number of levels.
Note that when $d = 1$, $(d,t)$-locating arrays are the same as $(\overline{d},t)$-locating arrays.

\subsection{Genetic Algorithms in CIT}

Metaheuristic algorithms have long played a role in the construction of combinatorial arrays; see Section~\ref{sec:related_work} for more details.
In this paper we employ a genetic algorithm, and provide an overview of them here. 
We form a \emph{population} $\mathcal{P}$, which is a set of \emph{individuals}.
Each individual is an array of a fixed size, but is not necessarily a covering or locating array; the number of columns in our setup will always be the number of system factors, but the (initial) number of rows will be determined after the first stage of our algorithm terminates (see Section~\ref{sec:the_two_stage_approach} for details).
Initially, each individual is generated uniformly-at-random (i.e., each entry is chosen uniformly at random, independently of other entries).
A fitness score is calculated for each individual based on the given problem; a common score for covering arrays is how many interactions are covered in the individual.
All the individuals are now ranked according to their score.
Determining the next population is based on \emph{selection}, \emph{crossover}, and \emph{mutation} operators.
Selection determines which individuals (``parents'') will produce ``offspring,'' crossover combines the ``genes'' of these parents to produce the offspring, and mutation randomly alters the offspring's genes.
The most fit individuals, among the original population and produced offspring, survive to the next generation. 
The process then repeats until a termination condition is reached.

\section{Approach}\label{sec:the_two_stage_approach}

This section contains our two-stage algorithm for constructing locating arrays. 
Locating arrays have two requirements: (1) all interactions must be covered at least a given number of times, and (2) all pairs of sets of interactions (at most a given size) must have sufficiently high symmetric difference on rows in which they appear. 
These requirements appear related but in some cases are not; covering arrays of course satisfy requirement (1), but make no guarantee about (2).
Suppose that $\mathcal{T}_1, \mathcal{T}_2$ are already found to be a locating pair within some covering array $\A$.
Then adding more rows to $\A$ will not cause them to become nonlocating. 
These facts together suggest a two-stage approach, where the first stage constructs a covering array, and the second stage appends rows to locate the remaining nonlocating pairs (if any exist).

Why bother with a two-stage approach, and not simply use an existing single method to directly build a locating array?
This is certainly possible, and there are a number of algorithms to do so \cite{seidel_separating_2018,lanus_partitioned_2019}; further, using a direct algorithm is arguably simpler and easier to write and maintain than coordinated multiple of them.
These methods suffer from one big disadvantage: determining nonlocating pairs is very intensive, both storage-wise (for some methods) and computationally (for all methods).
There are only a small number of system setups for which a direct locating array construction is known, thus computationally finding locating arrays is necessary for most cases.
We give our reasoning about computational cost through $(d,t)$-locating arrays for simplicity; the computation/storage costs are higher for $(\overline{d}, t)$-locating arrays.
The number of interactions for a covering array is $\binom{k}{t} v^t$, and thus the number of $d$-sets of interactions is $S = \binom{\binom{k}{t} v^t}{d}$.
The number of pairs of $d$-sets to check non-location is $P = \binom{S}{2}$; when $d > 1$ and other parameters are moderately large, $S$ is extremely large. 
Suppose that we have a system with $k = 20$ factors, $v = 3$, and strength $t = 2$. 
This system contains 1710 interactions. 
If $d = 1$, then $S = 1710$, and $P = 1461195$; if $d = 2$, then $S = 1461195$, and $P$ is greater than 1 trillion.
This is just to iterate through all pairs \emph{once}; metaheuristic approaches generally need to compute the number of nonlocating pairs many times.
Therefore, when $d > 1$, iterating through all pairs is largely infeasible; and even if $d = 1$, $P$ is still very large, even for moderately complex systems. 

We give further motivation for having two stages.
Empirically, many pairs of $d$-sets within a covering array are also locating, provided $d$ is sufficiently small.
Consider a row-optimal $\CA_1(7; 2, 15, 2)$ by the methods of Kleitman \cite{kleitman_families_1973} and Spencer \cite{spencer_minimal_1971}, and independently by Katona \cite{katona_two_1973}, reproduced in Figure~\ref{fig:ca_example_num_cov}.
Every interaction appears at least once, but there are many interactions appearing more than once. 
Indeed, if the number of rows is larger than $\lambda v^t$, such interactions always exist.
In the case of Figure~\ref{fig:ca_example_num_cov}, 165 interactions (39.3\%) are covered once, 195 (46.4\%) are covered twice, and 60 (14.3\%) are covered thrice.

\begin{figure}
    \centering
    \[ \begin{array}{ccccccccccccccc}
0 & 0 & 0 & 0 & 0 & 0 & 0 & 0 & 0 & 0 & 0 & 0 & 0 & 0 & 0\\
0 & 0 & 0 & 0 & 0 & 1 & 1 & 1 & 1 & 1 & 1 & 1 & 1 & 1 & 1\\
0 & 1 & 1 & 1 & 1 & 0 & 0 & 0 & 0 & 1 & 1 & 1 & 1 & 1 & 1\\
1 & 0 & 1 & 1 & 1 & 0 & 1 & 1 & 1 & 0 & 0 & 0 & 1 & 1 & 1\\
1 & 1 & 0 & 1 & 1 & 1 & 0 & 1 & 1 & 0 & 1 & 1 & 0 & 0 & 1\\
1 & 1 & 1 & 0 & 1 & 1 & 1 & 0 & 1 & 1 & 0 & 1 & 0 & 1 & 0\\
1 & 1 & 1 & 1 & 0 & 1 & 1 & 1 & 0 & 1 & 1 & 0 & 1 & 0 & 0
\end{array}\] 
% 000000000000000\\
% 000001111111111\\
% 011110000111111\\
% 101110111000111\\
% 110111011011001\\
% 111011101101010\\
% 111101110110100
    \caption{A row-optimal $\CA_1(7; 2, 15, 2)$ by the methods of Kleitman \cite{kleitman_families_1973} and Spencer \cite{spencer_minimal_1971}, and independently by Katona \cite{katona_two_1973}.}
    \label{fig:ca_example_num_cov}
\end{figure}

\begin{figure*}
    \centering
    \includegraphics[width=0.7\textwidth]{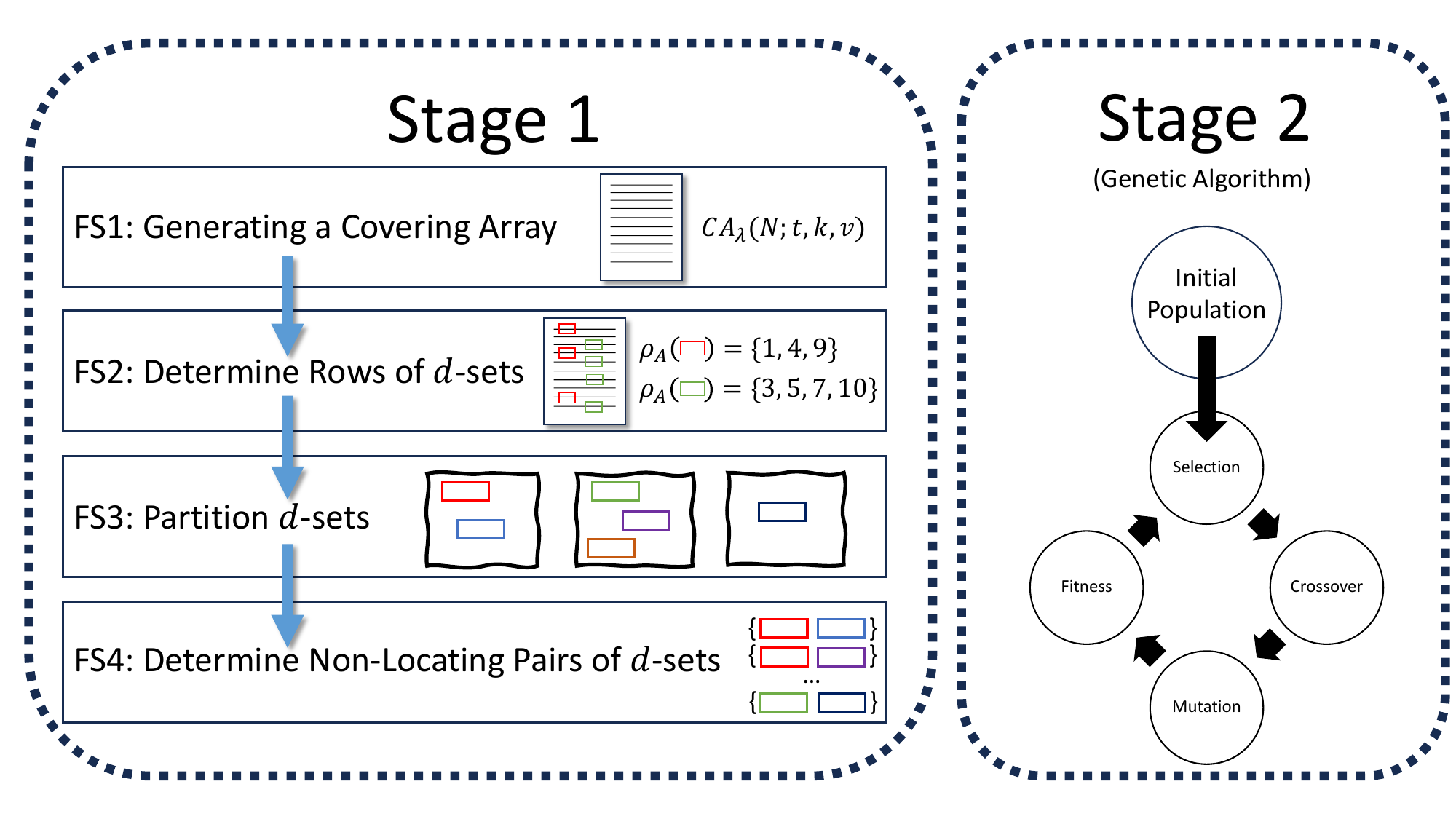}
    \caption{Main overview of \mainalgname's workflow. The first stage is divided into four sub-stages: (\firststagenum{1}) generating an initial covering array $\A$, (\firststagenum{2}) determining the rows of $\A$ for each $d$-set of interactions, (\firststagenum{3}) partitioning these $d$-sets based on their rows, and (\firststagenum{4}) using this partition to intelligently determine non-locating pairs.
    The second stage takes the output from the first stage (the non-locating pairs) and runs a genetic algorithm to find a small number of rows to add to $\A$ so that all non-locating pairs are now locating.
    }
    \label{fig:main_alg_overview}
\end{figure*}

In this section, we outline a two-stage approach, which we call \mainalgname, to generate locating arrays; the first is in Section~\ref{subsec:first_stage}, and the second in Section~\ref{subsec:second_stage}.
The overall goal of the first stage is to produce a covering array with redundancy $\lambda$ with few rows and determine all nonlocating pairs from the covering array.
The second stage produces as few rows as necessary to locate these remaining pairs, and employs a genetic algorithm.
The workflow of \mainalgname\ is shown in Figure~\ref{fig:main_alg_overview}.

\subsection{First Stage}\label{subsec:first_stage}

The first stage of \mainalgname\ has four sub-parts: (\firststagenum{1}) generation of an initial covering array $\A$, (\firststagenum{2}) determining the rows of $d$-sets within $\A$, (\firststagenum{3}) partitioning these $d$-sets into appropriate groups, and finally (\firststagenum{4}) determining non-locating pairs of $d$-sets based on this grouping. 

We begin with \firststagenum{1} about generating an initial covering array.
We use the CAGen tool \cite{wagner_cagen_2020}, which provides support for generating higher-index covering arrays.
The method is based on the in-parameter-order strategy \cite{lei_-parameter-order_1998}; we provide high-level pseudocode of their method in Algorithm~\ref{alg:cagen_outline}.
In general, any covering array generation method with support for higher indices would suffice, such as that of Dougherty et al. \cite{dougherty_algorithmic_2022}.
Even though CAGen uses a specific generation strategy, we do not make any effort to optimize our method by assuming a specific internal structure of the covering array, such as generation from some mathematical object \cite{meagher_group_2005,sherwood_covering_2006,colbourn_asymptotic_2018} or from a recursive construction \cite{colbourn_combinatorial_2004}.
We return to this as future work in Section~\ref{sec:conclusion}.

\begin{algorithm}
    \caption{\label{alg:cagen_outline}\firststagenum{1}: CAGen Overview (Initial $\CA$ Generation)}
    \begin{algorithmic} 
        \STATE $\A \gets$ exhaustive array on $t$ columns with index $\lambda$.
        \FOR{$i \gets t, \cdots, k$}
            \STATE Horizontally extend $\A$ by one column.
            \IF{$\A$ contains an uncovered interaction}
                \STATE Vertically extend $\A$ until all such interactions are $\lambda$-covered.
            \ENDIF
        \ENDFOR
        \RETURN $\A$.
    \end{algorithmic}
\end{algorithm}

For the second sub-part \firststagenum{2}, we determine the rows for all $d$-sets of interactions $D$, the process of which comes from the definition of locating arrays.
For each interaction $I$ in $D$, we determine $\rho_{\A}(I)$ and union all such sets of rows together for $D$.
Pseudocode is presented in Algorithm~\ref{alg:determine_rows_of_d_sets}.

\begin{algorithm}
    \caption{\label{alg:determine_rows_of_d_sets}\firststagenum{2}: Determining Rows of $\overline{d}$-Sets}
    \begin{algorithmic} 
        \STATE $\A \gets$ a covering array $\CA_\lambda(N; t, k, v)$ from Algorithm~\ref{alg:cagen_outline}.
        \FORALL{sets of interactions $D$ of size at most $d$}
            % \FORALL{interactions $I$ in $D$}
            %     \STATE $\rho_{\A}(I) \gets$ the set of rows in which $I$ appears in $\A$
            % \ENDFOR
            \STATE Set $\rho_{\A}(D) \gets \bigcup_{I \in D} \rho_{\A}(I)$
        \ENDFOR
        \RETURN $\rho_{\A}$.
    \end{algorithmic}
\end{algorithm}

The first non-trivial algorithm is for the third sub-part \firststagenum{3}, presented in Algorithm~\ref{alg:map_of_lamb_rows_to_d_sets}.
We create a data structure $RowMap$ consisting of pairs, where the first element is an integer, which represents how many rows in which a $d$-set appears, and the second element is a collection of $d$-sets and the rows in which these $d$-sets appear.
Since in \firststagenum{1} a covering array with redundancy $\lambda$ was created, every interaction must appear at least $\lambda$ times; however, some interactions may appear many more times. 
The meaning of $RowMap$ is to group $d$-sets into smaller collections based on the number of rows in which they appear. 
% Although we cannot guarantee that any interaction appears more than $\lambda$ times
Consider two elements $K_1, K_2$ in $RowMap$, and observe the first element $n_1, n_2$ of each.
If $n_1$ and $n_2$ are sufficiently different, then any pair of $d$-sets they represent (where one comes from $K_1$ and the other from $K_2$) appear that many times differently, which would imply these $d$-sets are a locating pair.
Therefore, we can avoid the computation of whether two $d$-sets are locating where one $d$-set comes from $K_1$, and the other from $K_2$.
Algorithm~\ref{alg:map_of_lamb_rows_to_d_sets} creates a dictionary with such a partition.

\begin{algorithm}
    \caption{\label{alg:map_of_lamb_rows_to_d_sets}\firststagenum{3}: Partitioning $\overline{d}$-Sets}
    \begin{algorithmic} 
        \STATE $\rho_{\A} \gets $ result of Algorithm~\ref{alg:determine_rows_of_d_sets}.
        \STATE $RowMap \gets$ empty dictionary, with keys being integers and values being arrays; insert keys 0 through $|\A|$ with empty initial values into $RowMap$.
        % with key being a set of $\lambda$ integers, and the value being a collection of tuples $(D, R)$ where $D$ is a $d$-set of interactions, and $R$ is a set of integers.
        \FORALL{sets of interactions $D$ of size at most $d$}
            \STATE $r \gets |\rho_{\A}(D)|$.
            \STATE Insert $(D, \rho_{\A}(D)))$ into $RowMap[r]$.
        \ENDFOR
        \RETURN $RowMap$.
    \end{algorithmic}
\end{algorithm}

For \firststagenum{4}, Algorithm~\ref{alg:determine_nonlocating} determines all nonlocating pairs. 
If two keys $K_1, K_2$ are equal, then we need to compare all distinct pairs of $d$-sets they represent; if not, then we compare all pairs.
In addition to each pair of $d$-sets added to $\nonlocDSname$, we also provide the size of the symmetric difference.
This will help improve the second stage of \mainalgname\ for the following reason.
Suppose that $D_1, D_2$ were found to be a nonlocating pair by Algorithm~\ref{alg:determine_nonlocating} with symmetric difference size $\ell$, and the goal is $\lambda$, where $\lambda > \ell$.
Then in the rows to-be-added in the second stage, $D_1, D_2$ only need to have a symmetric difference size $\lambda - \ell$.

\begin{algorithm}
    \caption{\label{alg:determine_nonlocating}\firststagenum{4}: Determine Non-Locating Pairs of $\overline{d}$-Sets}
    \begin{algorithmic} 
        \STATE $RowMap \gets$ result of Algorithm~\ref{alg:map_of_lamb_rows_to_d_sets}.
        \STATE $NonLoc \gets$ empty array.
        \FORALL{$\{(K_1, V_1), (K_2, V_2)\} \subseteq RowMap$}
            \IF{$|K_1 - K_2| \ge \lambda$}
                \STATE continue.
            \ENDIF
            \IF{$K_1 \ne K_2$}
                \FORALL{$(D_1, R_1) \in V_1, (D_2, R_2) \in V_2$}
                    \IF{the symmetric difference of $R_1, R_2$ is less than $\lambda$}
                        \STATE $\ell \gets$ size of this symmetric difference.
                        \STATE Add $(D_1, D_2, \ell)$ to $\nonlocDSname$.
                    \ENDIF 
                \ENDFOR
            \ELSE 
                \FORALL{$\{(D_1, R_1), (D_2, R_2)\} \subseteq V_1$}
                    \IF{the symmetric difference of $R_1, R_2$ is less than $\lambda$}
                        \STATE $\ell \gets$ size of this symmetric difference.
                        \STATE Add $(D_1, D_2, \ell)$ to $\nonlocDSname$.
                    \ENDIF 
                \ENDFOR
            \ENDIF
        \ENDFOR
        \RETURN $\nonlocDSname$.
    \end{algorithmic}
\end{algorithm}

The runtime of Algorithm~\ref{alg:determine_nonlocating} is not easily analyzed as that requires knowledge about how many times each interaction is covered a priori (as well as which $d$-sets have the same number of rows).
Regardless, Algorithm~\ref{alg:determine_nonlocating} has highest performance when (1) the number of $d$-sets within each value of $RowMap$ is relatively balanced, (2) the number of rows in $\A$ is large, and (3) $d$ is small.
For (1), in practice the sizes within $RowMap$ follow a ``bell curve'' shape.
Observe Table~\ref{tbl:rowmap_size_example} for an example of this behavior.
We used CAGen to generate a covering array $\CA_2(168; 3, 40, 3)$, and for $d=1$ took note of how many $d$-sets are covered each number of times.

\begin{table*}[]
\begin{tabular}{|l|l|l|l|l|l|l|l|l|l|l|l|l|l|l|l|}
\hline
$|\rho_{\A}(\mathcal{T})|$     & 2   & 3   & 4    & 5    & 6    & 7    & 8    & 9    & 10   & 11  & 12 & 13 & 14 & 15 & 16\\ \hline
\# $d$-Sets & 2454 & 10490 & 28681 & 51545 & 62414 & 52722 & 32967 & 15856 & 6356 & 2226 & 746 & 227 & 61 & 8 & 7 \\ \hline
\end{tabular}
\caption{\label{tbl:rowmap_size_example}For a $\CA_2(168; 3, 40, 3)$ generated by CAGen, we report the number of times any $d$-set is covered in the array as well as how many $d$-sets are covered that many times.}
\end{table*}

For (3), in all experiments we performed, we had $d \in \{1, 2\}$.
We address (2) next.
The CAGen tool produces small covering arrays quickly, but since Algorithm~\ref{alg:determine_nonlocating} is much slower on these small arrays than big ones, when $d > 1$, we also employ a different covering array generation approach for comparison.
Specifically, we use the Lov\'asz local lemma construction from Calbert and Dougherty \cite{calbert_upper_2023}; we denote this method as LLL.
This method constructs a uniformly random initial array (i.e., all entries chosen uniformly-at-random, independent of other entries) of a size that meets asymptotic growth, and checks if it is a covering array.
If so, then it terminates.
Otherwise, all columns of the first uncovered interaction's columns are resampled; the check is repeated until a covering array is found.
The number of rows LLL produces is higher than that of CAGen in practice, but is both asymptotically optimal and simpler to implement than CAGen.
Even though the number of rows LLL produces is higher than CAGen, there are both more partitions and fewer nonlocating pairs in general for the second stage. 
Very often the first stage with LLL will locate all pairs of $d$-sets.

\subsubsection{Optimizations}

We now describe additional optimizations to the above algorithms that, although do change the big-O runtime overall slightly, nevertheless improve performance in practice.
One big optimization is with using sorted arrays instead of sets, especially the second elements within each value of $RowMap$ during Algorithm~\ref{alg:map_of_lamb_rows_to_d_sets}.
This choice has two advantages.
The first is that computing the symmetric difference can be done in linear time in the size of the original collections by a simple ``incrementing'' algorithm.
The second is during both the inner and outer loops of Algorithm~\ref{alg:determine_nonlocating}.
One can envision each of these loops instead as a nesting of two loops.
Since we only need to consider 2 $d$-sets at a time, and the order does not matter, we iterate through $RowMap$ in sorted order for the outer loop, where the second loop for $(K_2, V_2)$ starts where $(K_1, V_1)$ is currently.
The behavior for the inner loops is similar.
Since we loop in sorted order, if the second set's smallest row number is larger than that of the first's, then we can immediately terminate the loop for the second.
The sorting happens in \firststagenum{3}; this sorting does increase the running time for this substage.
However, this increase is negligible overall as the vast majority of the runtime in the first stage is within \firststagenum{4}.

\subsection{Second Stage}\label{subsec:second_stage}

If there are no pairs in $\nonlocDSname$ from Algorithm~\ref{alg:determine_nonlocating}, then we skip the second stage; otherwise, we continue. 
The goal of \mainalgname's second stage is to add rows to an existing covering array $\A$ from the first stage that support location for all pairs in $\nonlocDSname$.
The number of rows needed will depend on the number of these pairs as well as what the pairs are; one could use a computationally expensive algorithm to do so, such as those developed by Seidel et al. \cite{seidel_separating_2018}.
To do so, we employ a genetic algorithm combined with binary search.
For background on the usage of genetic algorithms and other metaheuristic techniques with covering arrays, see Timan\'{a}-Pe\~{n}a et al. \cite{timana-pena_metaheuristic_2016}.

Individuals within our genetic algorithm are represented by an $N' \times k$ array ($N'$ rows, $k$ columns), with initially all entries set to a uniformly-at-random value between $0$ and $v-1$; we describe how to calculate $N'$ next.
Some existing genetic algorithm approaches \cite{stardom_metaheuristics_2001} require the number of rows a priori; we use a method inspired by that of Sabharwal et al. \cite{sabharwal_construction_2017}.
Suppose that we have determined that our GA was not able to find an individual that locates all pairs in $\nonlocDSname$ with $N_{lo}$ rows, but was able to do so with $N_{hi}$ rows, where $N_{hi} > N_{lo}$.
We then run binary search within the range $(N_{lo}, N_{hi}]$ to find as-small-as-possible number of rows to add $N'$ by repeatedly running our GA.
We keep track of the most-fit individuals throughout the process. 
To start this process, we initialize $N'$ to be the largest $\ell$ in $\nonlocDSname$, as that many rows are required (at least) to locate that pair.
Note that interactions are not required to appear (separate from location), as they are guaranteed to be covered by \firststagenum{1}.
While the GA fails, we multiply $N'$ by 2; this process terminates as the optimal value of $N'$ is at most the size of $\nonlocDSname$. 
When the GA finally succeeds with a value $N'$, we set $N_{hi} \gets N'$ and $N_{lo} \gets N' / 2$. 
Since locating array sizes grow logarithmically \cite{seidel_separating_2018,colbourn_combinatorial_2018}, one could set $N'$ initially to be a value that is guaranteed to work; this will lower the number of GA iterations slightly, but we do not pursue this here for simplicity.

Once the fitness of all individuals is calculated, we take the most fit half of the population, and discard the rest. 
We then select parents for crossover and/or mutation until we generate enough individuals to restore the population size. 
Mutation and crossover operators for the GA are standard.
Whenever mutation is considered, we choose a random number $r_{mut}$ between 1 and 3. 
If $r_{mut} = 1$, we randomly sample a new row and place it in a random position of the individual.
If $r_{mut} = 2$, we do the same but with an entire column.
And if $r_{mut} = 3$, we randomly change one entry.
For crossover, we generate a random number $r_{cross} \in \{1, 2\}$.
If $r_{cross} = 1$, we perform 1-point crossover in terms of the rows of the parents; otherwise, we perform 2-point crossover.
The fitness function determines how many locating pairs there are among those in $\nonlocDSname$ within the individual. 
Our GA has 100 individuals, runs for 100 generations, has mutation set at 30\%, crossover at 10\%, and employs uniformly-at-random selection of parents each generation.

\section{Experimental Design and Evaluation}\label{sec:experimental_evaluation}
\newcommand{\researchquestionnum}[1]{\textbf{RQ#1}}

We conducted all experiments in this paper on an Apple Silicon M2 Max processor with 64GB of memory.
Both stages of \mainalgname\ were written in C++.\footnote{For ease of generating interactions and $d$-sets of them in the first stage, we used the CppItertools project by Haining (\url{https://github.com/ryanhaining/cppitertools}) that generates combinations of a collection.}
For each parameter situation, we ran 10 iterations of the second stage when using CAGen (as it is deterministic and the GA is not), and 10 iterations of both the first and second stage when using the LLL method (as they are both randomized).
All times are for single-threaded operation, as that allows for a more direct comparison to existing methods.

The \emph{type} of a $\LA_\lambda(N; d, t, k, v)$ is denoted as $v^k$; for example, a test suite with 20 factors and 2 levels each has type $2^{20}$.
We consider different types in our evaluation of \mainalgname, namely when $v \in \{2, 3\}$ and $k \in \{10, 15, 20, 30, 40, 50, 100\}$ for $d=1$; and $v=3, 10 \le k \le 16$ for $d=2$.
We set a timeout of 1 hour (across both stages) for each type when $d=1$, and 2 hours when $d=2$.

We now investigate the effectiveness of \mainalgname\ for constructing locating arrays.  
We formulate the following research questions:
\begin{itemize}
    \item \researchquestionnum{1}: How much faster does \mainalgname\ produce locating arrays than existing methods for $d=1$, if at all?
    \item \researchquestionnum{2}: How much smaller are the locating arrays produced by \mainalgname\ compared to existing methods for $d=1$, if at all?
    \item \researchquestionnum{3}: Can \mainalgname\ produce locating arrays for $d > 1$ in a ``reasonable'' amount of time?
\end{itemize}

\subsection{\researchquestionnum{1}: Performance of \mainalgname}

We answer \researchquestionnum{1} from two perspectives: (1) a comparison between the number of iterations/found nonlocating pairs in the first stage of \mainalgname\ and that of existing approaches, and (2) overall time across both stages of \mainalgname.
% Note that for answering (2), the main possible performance improvements only come from the first stage.
% This is because the representation of individuals for the GA in the second stage is standard, apart from which pairs of $d$-sets it considers; however, these come from the first stage (and any existing GA method would similarly increase in performance).

\subsubsection{First Stage Iterations Comparison}

To see how much this first stage reduces the size of $\nonlocDSname$, we ran an experiment.
We set $d \in \{1, 2\}, t = 2, 10 \le k \le 16, v = 3$, and $\lambda = 2$, and for each scenario generated covering arrays using the LLL and CAGen methods.
As there may not be a locating array with $d \ge v$ \cite{colbourn_locating_2008}, this is the first computationally challenging case for $d=2$.
In Table~\ref{tbl:initial_experiment_speedup_show}, we show the overall total number of pairs of $d$-sets, the sizes of covering arrays produced by both the LLL and CAGen methods, and the number of $d$-sets compared within \firststagenum{4} in Algorithm~\ref{alg:determine_nonlocating} using these covering arrays. 
It is immediately apparent that the number of nonlocating pairs is indeed much smaller than the total number of pairs.
Additionally, the average number of found nonlocating pairs is many thousands of times smaller than the number of pairs. 
For example, for $d = 2, k = 15$, for CAGen the reduction is by a factor of over 20 thousand; for LLL it is even more dramatic, at 10 million times smaller (with a sacrifice in the number of rows).
This gives strong evidence that our method is significantly faster than a brute-force method for fitness calculation.

\begin{table}[]
\centering
\begin{tabular}{l||l|l|l|l|l|l|}
\hline
& & & \multicolumn{2}{c|}{LLL} & \multicolumn{2}{c|}{CAGen} \\ \cline{4-7}
 & $k$ & \# Pairs & $N$ & $|\nonlocDSname|$ & $N$ & $|\nonlocDSname|$ \\ \hline
\multirow{6}{*}{$d=1$} & 10  & 81810  & 71 & 10  &  24 & 621        \\ 
& 11  & 122265   & 72 & 0  &  33& 361       \\ 
&12  & 176121     & 73& 0 &  33& 519       \\ 
&13  & 246051    & 74& 1  & 33& 785          \\ 
&14  & 334971    & 75& 0 &  33& 914          \\ 
&15  & 446040    & 75& 2 &   33& 1057     \\ 
&16  & 582660   & 76& 1 &    33& 1299      \\ \hline
\multirow{6}{*}{$d=2$} & 10  & 3346397145  & 71& 2094  &  24& 2334285        \\ 
& 11  & 7474303980   & 72& 5765  &  33& 589839       \\ 
&12  & 15509215260     & 73& 4114  &  33& 1205422       \\ 
&13  & 30270424275    & 74& 4945  & 33& 2778153          \\ 
&14  & 56102617935    & 75& 6485 & 33 & 3489916           \\ 
&15  & 99475617780    & 75& 9458 & 33 & 4822635      \\ 
&16  & 169746046470   & 76 & 12291 & 33 &  7961747      \\ 
\end{tabular}
\caption{\label{tbl:initial_experiment_speedup_show}Comparison between LLL and CAGen for the number of nonlocating pairs after Algorithm~\ref{alg:determine_nonlocating} for $(1,2)$ and $(2,2)$-locating arrays. Here, $10 \le k \le 16, v = 3, \lambda = 2$. 
The table reports the total number of pairs for each situation, the initial covering array size for each method, and the number of nonlocating pairs for each. 
}
\end{table}

We analyze the memory overhead of our approach compared to a brute-force one, specificially for iterating over pairs of $d$-sets.
There are two brute-force approaches: calculating $\rho_{\A}(\mathcal{T})$ on-the-fly for each pair of $d$-sets (slower), and storing them in a data structure for lookup (faster).
Note that the faster brute-force method must store all $d$-sets of interactions with the rows that cover their interactions.
The main overhead with our approach is with the construction of $RowMap$.
We show that the number of keys in $RowMap$ is far smaller than the number of $d$-sets.
Consider the extreme case of every possible number of rows being represented in $RowMap$; there would be $N$ such sets. 
LLL will guarantee that $N$ is $O(v^t \log k + \lambda v^t)$ by the analysis of Calbert and Dougherty \cite{calbert_upper_2023}.
The number of interactions, not even considering $d$-sets of them, is already vastly larger than this. 
Therefore, our method is nearly identical in memory usage compared to the faster brute-force method.
Even though this memory overhead is much larger than for the slower brute-force method, it is many thousands of times faster (implied by Table~\ref{tbl:initial_experiment_speedup_show}) as such methods in general have to repeat the calculation many times. 

\subsubsection{Overall Runtime for Both Stages}

We give an analysis of the runtime of the first stage of \mainalgname; empirical results are in Subsection~\ref{subsec:results_sizes_of_las}.
The main advantage is a lower storage requirement than existing approaches, notably conditional expectation (CE) \cite{seidel_separating_2018}, a one-row-at-a-time approach.
There, all pairs of $d$-sets need to be stored to determine an appropriate next row that locates as many pairs as possible.
If there are $S$ $d$-sets (as calculated previously), then CE's storage requirement is $O(S^2)$.
Our approach only stores each $d$-set once, and the rows of the initial covering array.
The storage therefore is $O(S \cdot N)$ where $N$ is the size of this covering array; since $N$ will be $O(v^t \log k + \lambda v^t)$ by the analysis of Calbert and Dougherty \cite{calbert_upper_2023}, $N$ is much smaller than $S$, yielding a large improvement over CE.
Additionally, CE's runtime involves iterating through all pairs at least once, which is $O(S^2)$ again. 
Our method in \firststagenum{3} only sorts each observed $d$-set once, which gives $O(N \cdot S \log S)$ runtime overall (as it must iterate through the covering array's rows), which is a very large improvement.

The second stage is harder to analyze as that would require knowledge of both how many nonlocating pairs are left, and what those pairs are.
If a collection of nonlocating pairs are sufficiently ``conflicting'' (i.e., require many rows for all pairs to become locating), then that would force the GA to run for more iterations.
As implied by Table~\ref{tbl:initial_experiment_speedup_show}, it is empirically the case that the number of nonlocating pairs is dramatically smaller than the original count.
Suppose there are $m$ nonlocating pairs remaining for the second stage. 
A simple analysis of randomized constructions of locating arrays \cite{seidel_separating_2018} shows that the number of rows required to locate $m$ nonlocating pairs is $O(\log m)$, where all parameters other than the number of nonlocating pairs (i.e., $d$, $v$, $t$, $\lambda$) are fixed.
Therefore, the GA will take $O(\log \log m)$ overall runs to find a suitable number of rows to add, in expectation, as the GA also constructs arrays randomly (initially).
If the GA's population contains $|\mathcal{P}|$ individuals, then the second stage's total runtime is $O(|\mathcal{P}| \cdot m \log \log m)$.
This analysis implies that if $|\mathcal{P}|$ is small, the GA will terminate quickly; however, this choice in general increases $m$.

\subsection{\researchquestionnum{2}: Sizes of \LA s from \mainalgname}\label{subsec:results_sizes_of_las}

This section addresses \researchquestionnum{2}.
As implied by our analysis in Section~\ref{sec:the_two_stage_approach}, the main bottleneck in \mainalgname\ is the runtime. 
In Tables~\ref{tbl:results_d1_v2} and \ref{tbl:results_d1_v3}, we give the results from our experimental setup for \mainalgname.
They contain the results of generating $(d,t)$-locating arrays with \mainalgname\ for $d=1, v \in \{2,3\}, 2 \le t \le 5$, and $k \in \{10,15,20,30,40,50,100\}$.
For each parameter situation, we consider when the initial array is generated by CAGen, when it is generated by the LLL method, and the best known size in the literature (if there is one, we provide a reference).
We provide both the smallest number of rows achieved along with the corresponding time taken. 
\mainalgname\ has a time-out of one hour; when generating a locating array would take longer, we note that with $\dagger$.

As is evident, the best known locating arrays are smaller than our results other than for a handful of situations.
However, we are able to generate such arrays much more quickly than existing methods.
Further, we generated locating arrays for parameters not previously possible, e.g., $t \ge 4$.

\newcommand{\timedout}{$\dagger$}

% Format is CAGen/LLL/best

\begin{sidewaystable}
    
\begin{tabular}{ll|ll|ll|ll|ll|}
\cline{3-10}
                  &  & \multicolumn{2}{c|}{$t=2$} & \multicolumn{2}{c|}{$t=3$} & \multicolumn{2}{c|}{$t=4$} & \multicolumn{2}{c|}{$t=5$}      \\ \cline{2-10} 
\multicolumn{1}{l|}{}   & Type      & \multicolumn{1}{c}{$N$}   & Time   & \multicolumn{1}{c}{$N$}   & Time   & \multicolumn{1}{c}{$N$}   & Time   & \multicolumn{1}{c}{$N$}   & Time \\ \hline
\multicolumn{1}{l|}{\multirow{7}{*}{$\lambda=1$}} & $2^{10}$  &   14/22/11 \cite{konishi_using_2020} &   0.3/0.0/0.1  &   36/60/$\star$ &   16.8/0.0/$\star$  &  87/144/$\star$ &  40.9/0.1/$\star$  &  191/315/$\star$ &  202.2/0.2/$\star$  \\
\multicolumn{1}{l|}{}   & $2^{15}$  &   16/23/15 \cite{konishi_using_2020} &   2.5/0.0/3.2  &  47/67/$\star$ &  85.4/0.1/$\star$  &  123/170/$\star$ &  484.9/1.8/$\star$  &  294/391/$\star$ &  1795.6/43.7/$\star$  \\
\multicolumn{1}{l|}{}   & $2^{20}$  &   18/24/17 \cite{konishi_using_2020} &   6.0/0.3/5.4  &  59/72/$\star$ &  169.7/0.4/$\star$  &  \timedout/186/$\star$ &  \timedout/32.7/$\star$  &  \timedout/444/$\star$ &  \timedout/580.9/$\star$  \\
\multicolumn{1}{l|}{}   & $2^{30}$  &   21/26/26 \cite{seidel_separating_2018} &   28.1/0.1/$\star$  &  67/80/$\star$ &  410.9/2.6/$\star$  &  \timedout/\timedout/$\star$ &  \timedout/\timedout/$\star$  &  \timedout/\timedout/$\star$ &  \timedout/\timedout/$\star$  \\
\multicolumn{1}{l|}{}   & $2^{40}$  &   24/28/26 \cite{seidel_separating_2018} &   45.4/0.7/$\star$  &  \timedout/86/$\star$ &  \timedout/14.1/$\star$  &  \timedout/\timedout/$\star$ &  \timedout/\timedout/$\star$  &  \timedout/\timedout/$\star$ &  \timedout/\timedout/$\star$  \\
\multicolumn{1}{l|}{}   & $2^{50}$  &   26/29/26 \cite{seidel_separating_2018} &   107.1/1.7/$\star$  &  \timedout/91/$\star$ &  \timedout/56.1/$\star$  &  \timedout/\timedout/$\star$ &  \timedout/\timedout/$\star$  &  \timedout/\timedout/$\star$ &  \timedout/\timedout/$\star$  \\
\multicolumn{1}{l|}{}   & $2^{100}$ &   33/36/31 \cite{seidel_separating_2018} &   617.6/16.1/$\star$  &  \timedout/\timedout/$\star$ &  \timedout/\timedout/$\star$  &  \timedout/\timedout/$\star$ &  \timedout/\timedout/$\star$  &  \timedout/\timedout/$\star$ &  \timedout/\timedout/$\star$  \\ \hline
\multicolumn{1}{l|}{\multirow{7}{*}{$\lambda=2$}} & $2^{10}$  &   20/30/19 \cite{seidel_separating_2018} &    0.6/0.0/$\star$  &  54/78/$\star$ &  25.2/0.0/$\star$  &  120/183/$\star$ &  40.9/0.1/$\star$  &  277/401/$\star$ &  257.4/0.7/$\star$  \\
\multicolumn{1}{l|}{}   & $2^{15}$  &   23/31/22 \cite{seidel_separating_2018} &  2.7/0.0/$\star$  &  69/87/$\star$ &  99.0/0.3/$\star$  &  153/211/$\star$ &  338.0/6.3/$\star$  &  364/482/$\star$ &  1483.7/157.1/$\star$  \\
\multicolumn{1}{l|}{}   & $2^{20}$  &   25/33/26 \cite{seidel_separating_2018} &  9.4/0.1/$\star$  &  77/92/$\star$ &  234.7/0.6/$\star$  &  \timedout/229/$\star$ &  \timedout/104.6/$\star$  &  \timedout/532/$\star$ &  \timedout/1948.4/$\star$  \\
\multicolumn{1}{l|}{}   & $2^{30}$  &   30/34/33 \cite{seidel_separating_2018} &   26.9/0.2/$\star$  &  \timedout/99/$\star$ &  \timedout/7.1/$\star$  &  \timedout/\timedout/$\star$ &  \timedout/\timedout/$\star$  &  \timedout/\timedout/$\star$ &  \timedout/\timedout/$\star$  \\
\multicolumn{1}{l|}{}   & $2^{40}$  &   32/35/33 \cite{seidel_separating_2018} &   51.7/0.3/$\star$  &  \timedout/106/$\star$ &  \timedout/44.8/$\star$  &  \timedout/\timedout/$\star$ &  \timedout/\timedout/$\star$  &  \timedout/\timedout/$\star$ &  \timedout/\timedout/$\star$  \\
\multicolumn{1}{l|}{}   & $2^{50}$  &   35/38/33 \cite{seidel_separating_2018} &   110.5/1.6/$\star$  &  \timedout/112/$\star$ &  \timedout/174.3/$\star$  &  \timedout/\timedout/$\star$ &  \timedout/\timedout/$\star$  &  \timedout/\timedout/$\star$ &  \timedout/\timedout/$\star$  \\
\multicolumn{1}{l|}{}   & $2^{100}$ &   41/45/39 \cite{seidel_separating_2018} &   511.2/14.7/$\star$  &  \timedout/\timedout/$\star$ &  \timedout/\timedout/$\star$  &  \timedout/\timedout/$\star$ &  \timedout/\timedout/$\star$  &  \timedout/\timedout/$\star$ &  \timedout/\timedout/$\star$  \\ \hline
\multicolumn{1}{l|}{\multirow{7}{*}{$\lambda=3$}} & $2^{10}$  &   24/36/24 \cite{seidel_separating_2018} &  0.6/0.0/$\star$  &  67/94/$\star$ &  28.0/0.0/$\star$  &  142/217/$\star$ &  32.0/0.3/$\star$  &  325/471/$\star$ &  152.0/1.4/$\star$  \\
\multicolumn{1}{l|}{}   & $2^{15}$  &   29/38/29 \cite{seidel_separating_2018} &  4.7/0.0/$\star$  &  85/103/$\star$ &  143.6/0.6/$\star$  &  185/248/$\star$ &  417.5/14.8/$\star$  &  432/578/$\star$ &  1706.6/268.7/$\star$  \\
\multicolumn{1}{l|}{}   & $2^{20}$  &   32/40/31 \cite{seidel_separating_2018} &   4.9/0.0/$\star$  &  99/109/$\star$ &  407.1/1.1/$\star$  &  \timedout/266/$\star$ &  \timedout/183.7/$\star$  &  \timedout/\timedout/$\star$ &  \timedout/\timedout/$\star$  \\
\multicolumn{1}{l|}{}   & $2^{30}$  &   36/42/40 \cite{seidel_separating_2018} &   20.3/0.2/$\star$  &  \timedout/118/$\star$ &  \timedout/13.2/$\star$  &  \timedout/\timedout/$\star$ &  \timedout/\timedout/$\star$  &  \timedout/\timedout/$\star$ &  \timedout/\timedout/$\star$  \\
\multicolumn{1}{l|}{}   & $2^{40}$  &   39/45/40 \cite{seidel_separating_2018} &   50.4/1.1/$\star$  &  \timedout/127/$\star$ &  \timedout/315.6/$\star$  &  \timedout/\timedout/$\star$ &  \timedout/\timedout/$\star$  &  \timedout/\timedout/$\star$ &  \timedout/\timedout/$\star$  \\
\multicolumn{1}{l|}{}   & $2^{50}$  &   43/46/40 \cite{seidel_separating_2018} &   82.6/2.0/$\star$  &  \timedout/\timedout/$\star$ &  \timedout/\timedout/$\star$  &  \timedout/\timedout/$\star$ &  \timedout/\timedout/$\star$  &  \timedout/\timedout/$\star$ &  \timedout/\timedout/$\star$  \\
\multicolumn{1}{l|}{}   & $2^{100}$ &   50/54/46 \cite{seidel_separating_2018} &   347.4/14.9/$\star$  &  \timedout/\timedout/$\star$ &  \timedout/\timedout/$\star$  &  \timedout/\timedout/$\star$ &  \timedout/\timedout/$\star$  &  \timedout/\timedout/$\star$ &  \timedout/\timedout/$\star$  \\ \hline
\multicolumn{1}{l|}{\multirow{7}{*}{$\lambda=4$}} & $2^{10}$  &   31/43/30 \cite{seidel_separating_2018} &   0.6/0.0/$\star$  &  87/109/$\star$ &  21.6/0.0/$\star$  &  215/248/$\star$ &  51.0/0.4/$\star$  &  497/536/$\star$ &  141.4/1.8/$\star$  \\
\multicolumn{1}{l|}{}   & $2^{15}$  &   37/45/34 \cite{seidel_separating_2018} &   4.0/0.0/$\star$  &  115/118/$\star$ &  118.4/0.2/$\star$  &  256/283/$\star$ &  441.5/19.7/$\star$  &  558/625/$\star$ &  1592.3/169.9/$\star$  \\
\multicolumn{1}{l|}{}   & $2^{20}$  &   41/46/37 \cite{seidel_separating_2018} &   10.3/0.1/$\star$  &  128/124/$\star$ &  348.8/1.5/$\star$  &  \timedout/299/$\star$ &  \timedout/242.5/$\star$  &  \timedout/\timedout/$\star$ &  \timedout/\timedout/$\star$  \\
\multicolumn{1}{l|}{}   & $2^{30}$  &   47/48/47 \cite{seidel_separating_2018} &   23.8/0.1/$\star$  &  \timedout/132/$\star$ &  \timedout/17.7/$\star$  &  \timedout/\timedout/$\star$ &  \timedout/\timedout/$\star$  &  \timedout/\timedout/$\star$ &  \timedout/\timedout/$\star$  \\
\multicolumn{1}{l|}{}   & $2^{40}$  &   50/50/47 \cite{seidel_separating_2018} &   43.1/0.9/$\star$  &  \timedout/141/$\star$ &  \timedout/108.6/$\star$  &  \timedout/\timedout/$\star$ &  \timedout/\timedout/$\star$  &  \timedout/\timedout/$\star$ &  \timedout/\timedout/$\star$  \\
\multicolumn{1}{l|}{}   & $2^{50}$  &   53/56/47 \cite{seidel_separating_2018} &   72.9/5.2/$\star$  &  \timedout/145/$\star$ &  \timedout/448.6/$\star$  &  \timedout/\timedout/$\star$ &  \timedout/\timedout/$\star$  &  \timedout/\timedout/$\star$ &  \timedout/\timedout/$\star$  \\
\multicolumn{1}{l|}{}   & $2^{100}$ &   64/59/53 \cite{seidel_separating_2018} &   742.1/13.0/$\star$  &  \timedout/\timedout/$\star$ &  \timedout/\timedout/$\star$  &  \timedout/\timedout/$\star$ &  \timedout/\timedout/$\star$  &  \timedout/\timedout/$\star$ &  \timedout/\timedout/$\star$  \\ \hline
\end{tabular}

\caption{\label{tbl:results_d1_v2}Results for \mainalgname\ in finding $(\overline{1},t,\lambda)$-locating arrays where $2 \le t \le 5, k \in \{10, 15, 20, 30, 40, 50, 100\}, v = 2, 1 \le \lambda \le 4$.
There are three values of $N$ and time reported for each $t$; what is reported by \mainalgname\ where CAGen generated the initial covering array, the same where LLL generated the initial covering array, and the smallest known $N$ in the literature.
The corresponding time values are those from the same locating array method (i.e., not necessarily the smallest time across all literature), and are reported in seconds.
When no $N$ is known from existing work, or no time is provided from existing work, a $\star$ is given.
A reference is also provided when available.
In case of timeout for \mainalgname, $\dagger$ is given.
}
\end{sidewaystable}

% Format is CAGen/LLL/best
\begin{sidewaystable}
\begin{tabular}{ll|ll|ll|ll|ll|}
\cline{3-10}
                  &  & \multicolumn{2}{c|}{$t=2$} & \multicolumn{2}{c|}{$t=3$} & \multicolumn{2}{c|}{$t=4$} & \multicolumn{2}{c|}{$t=5$}      \\ \cline{2-10} 
\multicolumn{1}{l|}{}   & Type      & \multicolumn{1}{c}{$N$}   & Time   & \multicolumn{1}{c}{$N$}   & Time   & \multicolumn{1}{c}{$N$}   & Time   & \multicolumn{1}{c}{$N$}   & Time \\ \hline
\multicolumn{1}{l|}{\multirow{7}{*}{$\lambda=1$}} & $3^{10}$  &   35/55/29 \cite{konishi_using_2020} &   7.4/0.0/4.0  &   132/235/$\star$ &   52.7/0.0/$\star$  &  \timedout/862/$\star$ &  \timedout/0.7/$\star$  &  \timedout/2918/$\star$ &  \timedout/2.4/$\star$  \\
\multicolumn{1}{l|}{}   & $3^{15}$  &   42/59/40 \cite{seidel_separating_2018} &   10.8/0.0/$\star$  &  165/262/$\star$ &  221.4/0.3/$\star$  &  \timedout/997/$\star$ &  \timedout/28.7/$\star$  &  \timedout/\timedout/$\star$ &  \timedout/\timedout/$\star$  \\
\multicolumn{1}{l|}{}   & $3^{20}$  &   45/61/44 \cite{seidel_separating_2018} &   35.0/0.0/$\star$  &  191/280/$\star$ &  523.1/1.9/$\star$  &  \timedout/1083/$\star$ &  \timedout/365.7/$\star$  &  \timedout/\timedout/$\star$ &  \timedout/\timedout/$\star$  \\
\multicolumn{1}{l|}{}   & $3^{30}$  &   54/65/57 \cite{seidel_separating_2018} &   117.1/0.0/$\star$  &  \timedout/304/$\star$ &  \timedout/25.7/$\star$  &  \timedout/\timedout/$\star$ &  \timedout/\timedout/$\star$  &  \timedout/\timedout/$\star$ &  \timedout/\timedout/$\star$  \\
\multicolumn{1}{l|}{}   & $3^{40}$  &   58/68/57 \cite{seidel_separating_2018} &   247.6/0.1/$\star$  &  \timedout/320/$\star$ &  \timedout/154.9/$\star$  &  \timedout/\timedout/$\star$ &  \timedout/\timedout/$\star$  &  \timedout/\timedout/$\star$ &  \timedout/\timedout/$\star$  \\
\multicolumn{1}{l|}{}   & $3^{50}$  &   63/69/57 \cite{seidel_separating_2018} &   411.6/0.2/$\star$  &  \timedout/\timedout/$\star$ &  \timedout/\timedout/$\star$  &  \timedout/\timedout/$\star$ &  \timedout/\timedout/$\star$  &  \timedout/\timedout/$\star$ &  \timedout/\timedout/$\star$  \\
\multicolumn{1}{l|}{}   & $3^{100}$ &   76/77/67 \cite{seidel_separating_2018} &   1985.0/4.4/$\star$  &  \timedout/\timedout/$\star$ &  \timedout/\timedout/$\star$  &  \timedout/\timedout/$\star$ &  \timedout/\timedout/$\star$  &  \timedout/\timedout/$\star$ &  \timedout/\timedout/$\star$  \\ \hline
\multicolumn{1}{l|}{\multirow{7}{*}{$\lambda=2$}} & $3^{10}$  &   51/74/46 \cite{seidel_separating_2018} &    5.9/0.0/$\star$  &  187/302/$\star$ &  36.3/0.0/$\star$  &  \timedout/1077/$\star$ &  \timedout/2.1/$\star$  &  \timedout/3587/$\star$ &  \timedout/7.0/$\star$  \\
\multicolumn{1}{l|}{}   & $3^{15}$  &   57/78/52 \cite{seidel_separating_2018} &  8.7/0.0/$\star$  &  221/331/$\star$ &  197.1/1.0/$\star$  &  \timedout/1221/$\star$ &  \timedout/90.3/$\star$  &  \timedout/\timedout/$\star$ &  \timedout/\timedout/$\star$  \\
\multicolumn{1}{l|}{}   & $3^{20}$  &   62/81/57 \cite{seidel_separating_2018} &  18.1/0.0/$\star$  &  253/350/$\star$ &  607.5/6.2/$\star$  &  \timedout/1313/$\star$ &  \timedout/1152.9/$\star$  &  \timedout/\timedout/$\star$ &  \timedout/\timedout/$\star$  \\
\multicolumn{1}{l|}{}   & $3^{30}$  &   72/86/70 \cite{seidel_separating_2018} &   59.1/0.1/180  &  \timedout/376/$\star$ &  \timedout/85.8/$\star$  &  \timedout/\timedout/$\star$ &  \timedout/\timedout/$\star$  &  \timedout/\timedout/$\star$ &  \timedout/\timedout/$\star$  \\
\multicolumn{1}{l|}{}   & $3^{40}$  &   77/88/70 \cite{seidel_separating_2018} &   138.1/0.3/180  &  \timedout/\timedout/$\star$ &  \timedout/\timedout/$\star$  &  \timedout/\timedout/$\star$ &  \timedout/\timedout/$\star$  &  \timedout/\timedout/$\star$ &  \timedout/\timedout/$\star$  \\
\multicolumn{1}{l|}{}   & $3^{50}$  &   82/90/70 \cite{seidel_separating_2018} &   299.7/0.7/180  &  \timedout/\timedout/$\star$ &  \timedout/\timedout/$\star$  &  \timedout/\timedout/$\star$ &  \timedout/\timedout/$\star$  &  \timedout/\timedout/$\star$ &  \timedout/\timedout/$\star$  \\
\multicolumn{1}{l|}{}   & $3^{100}$ &   98/97/81 \cite{seidel_separating_2018} &   1752.5/12.5/$\star$  &  \timedout/\timedout/$\star$ &  \timedout/\timedout/$\star$  &  \timedout/\timedout/$\star$ &  \timedout/\timedout/$\star$  &  \timedout/\timedout/$\star$ &  \timedout/\timedout/$\star$  \\ \hline
\multicolumn{1}{l|}{\multirow{7}{*}{$\lambda=3$}} & $3^{10}$  &   64/91/57 \cite{seidel_separating_2018} &  6.7/0.0/$\star$  & 264/359/$\star$ &  41.7/0.1/$\star$  &  \timedout/1260/$\star$ &  \timedout/3.5/$\star$  &  \timedout/4157/$\star$ &  \timedout/9.2/$\star$  \\
\multicolumn{1}{l|}{}   & $3^{15}$  &   70/96/65 \cite{seidel_separating_2018} &  27.5/0.0/$\star$  &  299/390/$\star$ &  273.5/1.8/$\star$  &  \timedout/1412/$\star$ &  \timedout/165.3/$\star$  &  \timedout/\timedout/$\star$ &  \timedout/\timedout/$\star$  \\
\multicolumn{1}{l|}{}   & $3^{20}$  &   90/99/69 \cite{seidel_separating_2018} &   48.4/0.0/$\star$  &  325/411/$\star$ &  412.8/10.7/$\star$  &  \timedout/\timedout/$\star$ &  \timedout/\timedout/$\star$  &  \timedout/\timedout/$\star$ &  \timedout/\timedout/$\star$  \\
\multicolumn{1}{l|}{}   & $3^{30}$  &   97/103/83 \cite{seidel_separating_2018} &   166.2/0.2/$\star$  &  \timedout/438/$\star$ &  \timedout/143.7/$\star$  &  \timedout/\timedout/$\star$ &  \timedout/\timedout/$\star$  &  \timedout/\timedout/$\star$ &  \timedout/\timedout/$\star$  \\
\multicolumn{1}{l|}{}   & $3^{40}$  &  100/106/83 \cite{seidel_separating_2018} &   327.8/0.5/$\star$  &  \timedout/\timedout/$\star$ &  \timedout/\timedout/$\star$  &  \timedout/\timedout/$\star$ &  \timedout/\timedout/$\star$  &  \timedout/\timedout/$\star$ &  \timedout/\timedout/$\star$  \\
\multicolumn{1}{l|}{}   & $3^{50}$  &   102/109/83 \cite{seidel_separating_2018} &   491.2/1.2/$\star$  &  \timedout/\timedout/$\star$ &  \timedout/\timedout/$\star$  &  \timedout/\timedout/$\star$ &  \timedout/\timedout/$\star$  &  \timedout/\timedout/$\star$ &  \timedout/\timedout/$\star$  \\
\multicolumn{1}{l|}{}   & $3^{100}$ &   \timedout/116/94 \cite{seidel_separating_2018} &   \timedout/22.0/$\star$  &  \timedout/\timedout/$\star$ &  \timedout/\timedout/$\star$  &  \timedout/\timedout/$\star$ &  \timedout/\timedout/$\star$  &  \timedout/\timedout/$\star$ &  \timedout/\timedout/$\star$  \\ \hline
\multicolumn{1}{l|}{\multirow{7}{*}{$\lambda=4$}} & $3^{10}$  &   86/106/66 \cite{seidel_separating_2018} &  8.3/0.0/$\star$  &  267/411/$\star$ &  49.9/0.2/$\star$  &  \timedout/1427/$\star$ &  \timedout/5.3/$\star$  &  \timedout/4676/$\star$ &  \timedout/14.7/$\star$  \\
\multicolumn{1}{l|}{}   & $3^{15}$  &   79/111/73 \cite{seidel_separating_2018} &   136.8/0.0/$\star$  &  326/444/$\star$ &  248.7/2.4/$\star$  &  \timedout/1585/$\star$ &  \timedout/225.5/$\star$  &  \timedout/\timedout/$\star$ &  \timedout/\timedout/$\star$  \\
\multicolumn{1}{l|}{}   & $3^{20}$  &   99/115/79 \cite{seidel_separating_2018} &   49.0/0.0/$\star$  &  343/465/$\star$ &  623.9/16.9/$\star$  &  \timedout/\timedout/$\star$ &  \timedout/\timedout/$\star$  &  \timedout/\timedout/$\star$ &  \timedout/\timedout/$\star$  \\
\multicolumn{1}{l|}{}   & $3^{30}$  &   110/119/95 \cite{seidel_separating_2018} &   77.3/0.2/$\star$  &  \timedout/494/$\star$ &  \timedout/213.8/$\star$  &  \timedout/\timedout/$\star$ &  \timedout/\timedout/$\star$  &  \timedout/\timedout/$\star$ &  \timedout/\timedout/$\star$  \\
\multicolumn{1}{l|}{}   & $3^{40}$  &   117/123/95 \cite{seidel_separating_2018} &   158.2/0.7/$\star$  &  \timedout/\timedout/$\star$ &  \timedout/\timedout/$\star$  &  \timedout/\timedout/$\star$ &  \timedout/\timedout/$\star$  &  \timedout/\timedout/$\star$ &  \timedout/\timedout/$\star$  \\
\multicolumn{1}{l|}{}   & $3^{50}$  &   122/125/95 \cite{seidel_separating_2018} &   506.9/1.9/$\star$  &  \timedout/\timedout/$\star$ &  \timedout/\timedout/$\star$  &  \timedout/\timedout/$\star$ &  \timedout/\timedout/$\star$  &  \timedout/\timedout/$\star$ &  \timedout/\timedout/$\star$  \\
\multicolumn{1}{l|}{}   & $3^{100}$ &   \timedout/133/107 \cite{seidel_separating_2018} &   \timedout/32.5/$\star$  &  \timedout/\timedout/$\star$ &  \timedout/\timedout/$\star$  &  \timedout/\timedout/$\star$ &  \timedout/\timedout/$\star$  &  \timedout/\timedout/$\star$ &  \timedout/\timedout/$\star$  \\ \hline
\end{tabular}

\caption{\label{tbl:results_d1_v3}Results for \mainalgname\ in finding $(\overline{1},t,\lambda)$-locating arrays where $2 \le t \le 5, k \in \{10, 15, 20, 30, 40, 50, 100\}, v = 3, 1 \le \lambda \le 4$.
There are three values of $N$ and time reported for each $t$; what is reported by \mainalgname\ where CAGen generated the initial covering array, the same where LLL generated the initial covering array, and the smallest known $N$ in the literature.
The corresponding time values are those from the same locating array method (i.e., not necessarily the smallest time across all literature), and are reported in seconds.
When no $N$ is known from existing work, or no time is provided from existing work, a $\star$ is given.
A reference is also provided when available.
In case of timeout for \mainalgname (1 hour), $\dagger$ is given.
}
\end{sidewaystable}

% Format is CAGen/LLL/best
\begin{table}[]
\centering
\begin{tabular}{ll|ll|}
\cline{3-4}
                  &  & \multicolumn{2}{c|}{$t=2$}       \\ \cline{2-4} 
\multicolumn{1}{l|}{}   & Type      & \multicolumn{1}{c}{$N$}   & Time  \\ \hline
\multicolumn{1}{l|}{\multirow{6}{*}{$\lambda=1$}} & $3^{5}$  & 49 & 0.3   \\
\multicolumn{1}{l|}{}   & $3^{6}$  &  56  &  1.6   \\
\multicolumn{1}{l|}{}   & $3^{7}$  &  59 & 5.7 \\
\multicolumn{1}{l|}{}   & $3^{8}$  &  61 & 18.5 \\
\multicolumn{1}{l|}{}   & $3^{9}$  &  67 &  47.4  \\
\multicolumn{1}{l|}{}   & $3^{10}$ &  72 &  112.8   \\ \hline
\multicolumn{1}{l|}{\multirow{6}{*}{$\lambda=2$}} & $3^{5}$  & 68 & 1.1  \\
\multicolumn{1}{l|}{}   & $3^{6}$  &  73  & 4.5   \\
\multicolumn{1}{l|}{}   & $3^{7}$  &  79 & 19.0 \\
\multicolumn{1}{l|}{}   & $3^{8}$  &  86 & 44.9 \\
\multicolumn{1}{l|}{}   & $3^{9}$  &  88 & 158.1  \\
\multicolumn{1}{l|}{}   & $3^{10}$ &  95 & 358.2  \\ \hline
\multicolumn{1}{l|}{\multirow{6}{*}{$\lambda=3$}} & $3^{5}$  & 84 & 1.7   \\
\multicolumn{1}{l|}{}   & $3^{6}$  &  93  & 7.3   \\
\multicolumn{1}{l|}{}   & $3^{7}$  &  95 & 19.0 \\
\multicolumn{1}{l|}{}   & $3^{8}$  &  105 & 95.8 \\
\multicolumn{1}{l|}{}   & $3^{9}$  &  110 & 268.3  \\
\multicolumn{1}{l|}{}   & $3^{10}$ &  111  & 3061.4    \\ \hline
\multicolumn{1}{l|}{\multirow{6}{*}{$\lambda=4$}} & $3^{5}$  & 99 & 2.7  \\
\multicolumn{1}{l|}{}   & $3^{6}$  &  107  &  12.3   \\
\multicolumn{1}{l|}{}   & $3^{7}$  &  113 & 48.0 \\
\multicolumn{1}{l|}{}   & $3^{8}$  &  121 & 141.7 \\
\multicolumn{1}{l|}{}   & $3^{9}$  &  126 & 1212.4  \\
\multicolumn{1}{l|}{}   & $3^{10}$ &  136  & 6993.0    \\ \hline
\end{tabular}
\caption{\label{tbl:results_d2_v3}Results for \mainalgname\ in finding $(\overline{2},t,\lambda)$-locating arrays where $t = 2, 5 \le k \le 10, v = 3, 1 \le \lambda \le 4$.
We used the LLL method for initial covering array generation.
The corresponding times are for both stages, and are reported in seconds.
}
\end{table}

\subsection{\researchquestionnum{3}: Generating $\LA$s with $d > 1$}\label{subsec:generating_la_d_bigger_1}

This subsection concerns answering \researchquestionnum{3}.
In Table~\ref{tbl:results_d2_v3}, we show the results of \mainalgname\ for generating $(\overline{2}, 2)$-locating arrays with $5 \le k \le 10, 1 \le \lambda \le 4$, and $v=3$.
This is the first nontrivial case as there may not exist locating arrays with $\overline{d} = \overline{2}$ and $v = 2$ \cite{colbourn_locating_2008}.
Here, we use the LLL method for initial array generation as CAGen's usage yields many more nonlocating pairs, causing the second stage to become much slower. 
The timeout for these experiments was 2 hours.
 
These are the first locating arrays with $\overline{d}, t$ both at least 2; even so, the time taken increases rapidly as parameters increase.
To consider the cases when $d$ and $t$ are larger, or even if $k > 10$ or $v > 3$, would require a new method or assumption about the underlying $d$-sets. 
Nevertheless, for ``small'' systems, \mainalgname\ can indeed generate locating arrays with $d > 1$ in a ``reasonable'' amount of time.

\section{Threats to Validity}\label{sec:threats}

We are aware of several threats to the validity of our approach.
The main threat involves the correctness of the written programs for both stages of \mainalgname.
First, to determine the correctness of the CAGen program for the first stage, we wrote an independent covering array verifier.
Next, we ran the program on the resulting initial covering arrays; all these covering arrays passed verification.
Second, for both stages we created implementations of the algorithms in Section~\ref{sec:the_two_stage_approach} in both Python and C++.
For the first stage and for each parameter situation in Subsection~\ref{subsec:results_sizes_of_las}, we fixed a single covering array and verified that the resulting $\nonlocDSname$ set is the same for both implementations.

There is a potential threat for the second stage of \mainalgname\ as genetic algorithms have not been applied to locating arrays as far as we are aware.  
We do not see any threat with the chosen representation, selection operators, and fitness function of our GA as they are standard with respect to covering arrays; see Stardom \cite{stardom_metaheuristics_2001} for details.

% Another threat to validity is ...

\section{Related Work}\label{sec:related_work}

Related works fall into four main categories: mathematical constructions, computational approaches, metaheuristic methods, and applications of locating arrays in experiments.
There are several works that give a general background for CIT, notably those by Kuhn et al. \cite{kuhn_introduction_2013} and Nie and Leung \cite{nie_survey_2011}.
Additionally there are many other covering array construction approaches, such as AETG \cite{cohen_aetg_1997} and the density algorithm \cite{bryce_density_2007}.

Locating arrays were first introduced by Colbourn and McClary \cite{colbourn_locating_2008} as a variant of the existing covering arrays, as well as a related type called detecting arrays; see \cite{colbourn_coverage_2016} for a more recent summary.
The algorithm to recover the significant interactions for a detecting arrays is more efficient than the same for a locating arrays, but their sizes are in general larger.
We do not pursue detecting arrays in our work, and leave such application for future work. 
There has been some work in finding optimal locating arrays for $d=1$ and $t=1$ \cite{tang_optimality_2012,colbourn_disjoint_2016}, with bounds for this case for detecting arrays \cite{colbourn_detecting_2019}.
Asymptotic bounds are pursued by Colbourn and Syrotiuk \cite{colbourn_combinatorial_2018}.
Some recursive constructions were proved by Colbourn and Fan \cite{colbourn_locating_2016}, although only for the $d=1$ case.
Very recently, Colbourn and Syrotiuk \cite{colbourn_detecting_2022} construct detecting arrays for $t=1$ and arbitrary $d$ using a variant of separating hash families. 
Detecting arrays have a stronger condition than for locating arrays, but contain more rows in general.
The produced arrays by many of these methods have some internal structure, and thus the methods in this paper could in principle reduce the number of to-be-compared $d$-sets.

Computational methods based on the Stein-Lov\'asz-Johnson paradigm \cite{stein_two_1974,lovasz_ratio_1975,johnson_approximation_1974} were pursued by Seidel et al. \cite{seidel_separating_2018}.
Nagamoto et al. \cite{nagamoto_locating_2014} developed a two-stage method for some locating array parameters that is similar in spirit to our method. 
Our work could be considered an extension based on the partitioning strategy we developed. 
Lanus et al. \cite{lanus_partitioned_2019} created a tree-based verification algorithm for locating arrays and was the first paper (as far as we are aware) to computationally create locating arrays with both $d > 1$; however, they only concerned \emph{main effects}, where $t = 1$ (see Colbourn and Syrotiuk for discussion on these effects in CIT \cite{colbourn_detecting_2019}).
There may potentially be some more optimizations in our approach based on theirs. 

Dougherty \cite{dougherty_genetic_2020} partitioned covering array generation into multiple stages (instead of just 1 or 2) using a genetic algorithm \cite{dougherty_genetic_2020}. Instead of using the algorithm to modify the arrays directly, the genetic algorithm helped determine the optimal number of stages in which to have every interaction covered $\lambda$ times.
In principle his methods can also be applied to locating array generation, as follows.
Suppose the pairs of $d$-sets to check are $\mathcal{P} = \{P_1, \cdots, P_s\}$.
Instead of having a large storage requirement as our work does here, determine an efficiently-computable partition of $\mathcal{P}$, construct locating arrays for each part, and vertically juxtapose all such arrays. 
Such methods were used for covering array generation using groups \cite{sarkar_covering_2016}.

Hnich et al. \cite{hnich_constraint_2006} used constraint-based methods for covering arrays, and this was later applied to finding $(\overline{1},2)$ and $(\overline{1},3)$-locating arrays by Konishi et al. \cite{konishi_finding_2020}.
Jin et al. \cite{jin_satisfiability-based_2018} used Satisfiability Modulo Theories for finding ``constrained'' $(1,2)$-locating arrays. 
The search space increases dramatically as $d$ and $t$ increase, as discussed in Section~\ref{sec:the_two_stage_approach}. 
Thus a new idea is required if constraint/satisfiability methods are to be employed for larger $d,t$. 
Jin, Shi, and Tsuchiya \cite{jin_constrained_2023} also used a satisfiability approach for finding detecting arrays where there is a constraint on the $d$-sets considered for $d \in \{1,2\}$, which involves a notably smaller set of $d$-sets compared to all $d$-sets.
Our approach is only for unconstrained locating arrays; we plan on incorporating constraints into our algorithm as future work.

Simple hill-climbing approaches often lead to local optima for covering array generation \cite{stardom_metaheuristics_2001}.
As a result, evolutionary computation has been employed to covering arrays in many forms, of which we name a few: simulated annealing \cite{torres-jimenez_simulated_2010,guo_effective_2018,torres-jimenez_new_2022}, genetic algorithms \cite{stardom_metaheuristics_2001,sabharwal_construction_2017,dougherty_genetic_2020}, tabu search \cite{nurmela_upper_2004,walker_ii_tabu_2009}, particle swarm optimization \cite{li_effective_2022}, and ant colony optimization \cite{chen_variable_2009}.
As far as we are aware, there has only been one previous work, by Konishi et al. \cite{konishi_using_2020}, that used evolutionary computation for finding locating arrays, specifically with simulated annealing.
To do so, they used a two-phase algorithm which continually binary searches between pre-calculated bounds until a locating array is found; this is the same idea as our second stage.
The second phase occurs as simulated annealing is used once again to refine the solution. This work indicated that simulated annealing can scale with large systems if $t = 2$, yet often is unable to produce locating arrays for larger $t$ values.

There have been several real-world experiments performed to identify significant interactions in screening experiments using locating arrays. 
Aldaco et al. \cite{aldaco_locating_2015} constructed a $(\overline{1}, \overline{2}, 2)$-locating array on a system with 75 factors for screening the response of TCP throughput.
Their approach was initially similar to ours: they first generated a covering array, and applied post-optimization \cite{nayeri_randomized_2013} to reduce its size; see Section~\ref{sec:future_work} for details on post-optimization.
Compton et al. \cite{compton_screening_2016} constructed a $(\overline{1}, \overline{2})$-locating array for 24 factors in screening a wireless testbed system.
Seidel et al. \cite{seidel_analysis_2018} analyzed the same testbed system, along with another larger wireless network system, with a ``heavy hitters'' tree search algorithm for more robust analysis of the significant interactions. 
Seidel et al. \cite{seidel_robustness_2019} analyze robustness of recovery as a function of environment nondeterminism. 
Very recently, Akhtar et al. \cite{akhtar_scalable_2023} further discuss the same methods with other real-world experiments and data sets.
Nearly all of the locating arrays used within screening experiments are locating arrays with $d=1$ and $t=2$, and thus can only recover one significant interaction. 
We hope that faster construction methods, such as that discussed in this paper, can aid future experiment analysis methods to more efficiently identify more than one such interaction. 

\section{Future Work}\label{sec:future_work}

As mentioned previously, detecting arrays in general employ more rows than locating arrays do, but the algorithm for recovery of significant interactions is much more efficient. 
We plan on both incorporating a version of \mainalgname\ that includes detecting arrays, as well as locating/detecting arrays of ``mixed'' types.
This last point is important as real-world systems do not necessarily have the same number of levels for each factor.

For the first stage of \mainalgname, the CAGen tool constructs covering arrays using the ``in-parameter-order'' algorithm \cite{lei_-parameter-order_1998}, which adds one column at a time and any necessary rows after the new column is added.
One could check $d$-sets of interactions in a similar way; if $\mathcal{T}_1, \mathcal{T}_2$ are already a locating pair, then adding rows will not make them nonlocating.
Additionally, the LLL method often constructed an array that is already locating, thus removing the need for the second stage. 
This gives a potential research direction of adding an intermediary stage based on the work of Nayeri et al. \cite{nayeri_randomized_2013} called ``postopimization.'' 
Applied to our setting, this method takes a locating array, rearranges rows, and randomizes ``don't care'' values until a row is determined to be redundant (which is then removed). 
This process repeats until a termination condition is reached. 

The second stage of \mainalgname, even though more efficient than a brute-force search as it only operates on a much smaller number of $d$-sets, was nevertheless often the slower of the two stages, especially when the other parameters were small enough.
An explanation is that most individuals in the population have very high fitness, but not what is needed to locate; this was observed for a related object by Dougherty \cite{dougherty_hash_2019}.
A multi-stage approach would be more suitable, such as stopping the GA when a target fitness value smaller than $|\nonlocDSname|$ is achieved, and starting another GA on the remaining pairs.
We are currently pursuing this direction.

\section{Conclusion}\label{sec:conclusion}

In this paper we introduced \mainalgname, a two-stage algorithm for constructing locating arrays that uses a simple but powerful partitioning strategy in its first stage, and a genetic algorithm in the second stage.
This strategy utilized the empirical observation that in nearly every covering array with redundancy $\lambda$, there are many interactions covered more than $\lambda$ times.
If there is sufficient difference between the number of times two sets of interactions are covered, then they automatically are locating pairs. 
The partitioning drastically reduces the number of pairs needed to be compared in determining whether or not they are locating. 
As we have shown, the number of pairs is much smaller than the total number of pairs; we then utilize a simple genetic algorithm to locate all the remaining pairs. 

As a result, we presented the first nontrivial locating arrays with (1) $t \in \{4, 5\}$, $v \in \{2, 3\}$ and $d=1$, and (2) $t = 2$ and $\overline{d} = \overline{2}$.
Further, the parameters for which a locating array already exists are handled much more quickly by our two-stage algorithm than previous approaches. 

\section*{Acknowledgments}

The opinions in the work are solely of the authors, and do not reflect those of the U.S. Army, U.S. Army Research Labs, the U.S. Military Academy, or the Department of Defense.

%%
%% The next two lines define the bibliography style to be used, and
%% the bibliography file.
\bibliographystyle{plain}
\bibliography{references}

\end{document}